\documentclass{Interspeech}

\interspeechcameraready

\title{SAKURA: On the Multi-hop Reasoning of Large Audio-Language Models Based on Speech and Audio Information}

\author[]{Chih-Kai}{Yang}
\author[equalcontribution]{Neo}{Ho}
\author[equalcontribution]{Yen-Ting}{Piao}
\author[]{Hung-yi}{Lee}

\affiliation[nocounter]{}{National Taiwan University}{Taiwan}

\email{chihkaiyang1124@gmail.com, hungyilee@ntu.edu.tw}
\keywords{Large audio-language model, multi-hop reasoning, benchmark}

\usepackage{comment}
\usepackage{cite}
\usepackage{hyperref}
\usepackage{url}
\usepackage{makecell}
\usepackage{multirow}
\usepackage{xcolor}
\usepackage{relsize}

\usepackage{svg}

\begin{document}

\maketitle

\begin{abstract}

Large audio-language models (LALMs) extend the large language models with multimodal understanding in speech, audio, etc. While their performances on speech and audio-processing tasks are extensively studied, their reasoning abilities remain underexplored. Particularly, their multi-hop reasoning, the ability to recall and integrate multiple facts, lacks systematic evaluation. Existing benchmarks focus on general speech and audio-processing tasks, conversational abilities, and fairness, but overlook this aspect. To bridge this gap, we introduce SAKURA, a benchmark assessing LALMs’ multi-hop reasoning based on speech and audio information. Results show that LALMs struggle to integrate speech/audio representations for multi-hop reasoning, even when they extract the relevant information correctly, highlighting a fundamental challenge in multimodal reasoning. Our findings expose a critical limitation in LALMs, offering insights and resources for future research.

\end{abstract}

\section{Introduction}

\begin{figure*}[t]
    \centering
    


    \includegraphics[width=0.9\linewidth]{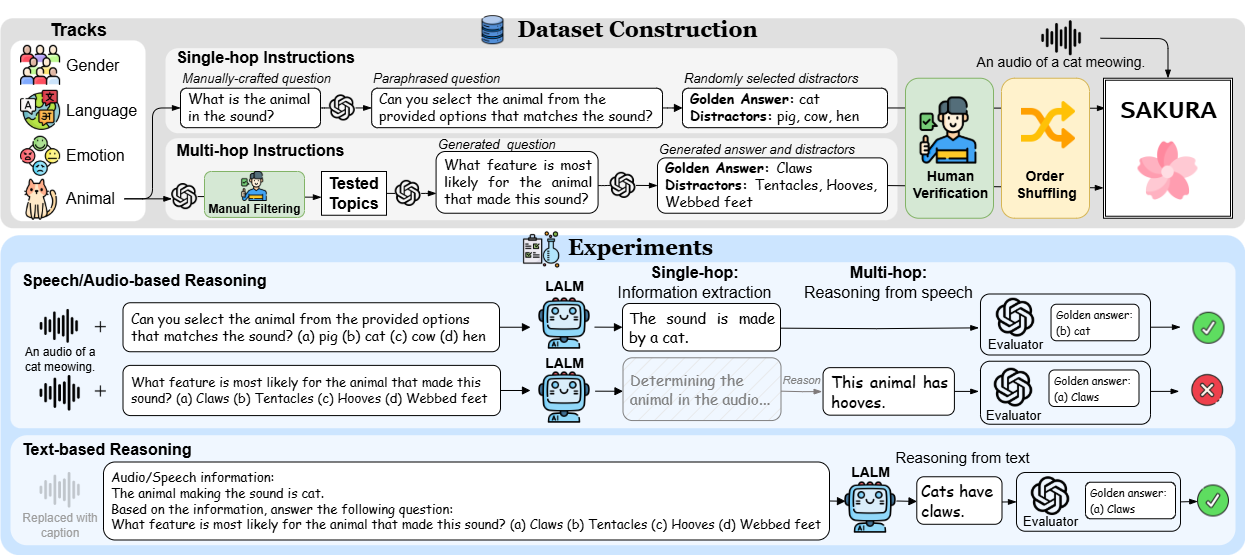}
    
    \vspace{-7pt}
    \caption{Overview of dataset construction and experiments. Question-answer pairs were generated by GPT-4o and manually verified. LALMs were evaluated on speech/audio-based vs. text-based reasoning, where text-based replaces auditory inputs with captions, assessing their ability to integrate auditory information. Icon source: \url{https://www.flaticon.com}}
    \label{fig:overview}

\vspace{-10pt}
\end{figure*}

Large language models (LLMs)\cite{llama31, gpt4o} have revolutionized AI research, extending beyond natural language processing to domain-specific applications like computer vision\cite{vipergpt} and speech processing~\cite{huang2023audiogpt, kuan2024speech}. This shift drives the rise of large multimodal models, e.g., large vision-language models (LVLMs)\cite{llava, gemini} and large audio-language models (LALMs)\cite{gong_ltuas, gama, salmonn, desta2, Qwen-Audio, Qwen2-Audio, gemini}, equipping LLMs with cross-modal understanding.

Research has also explored LLMs' fundamental reasoning abilities~\cite{jiang-etal-2024-peek, huang2024large, chen2024premise}, with \textbf{multi-hop reasoning}, the skill of recalling and connecting multiple facts, recognized as a key to complex reasoning~\cite{press-etal-2023-measuring, shalev2024distributional, latentmultihop, biran-etal-2024-hopping}. Particularly, latent multi-hop reasoning~\cite{latentmultihop, biran-etal-2024-hopping}, which involves multi-hop reasoning over internal parameterized knowledge stored in LLMs, is crucial for efficiently utilizing learned knowledge without heavily relying on external retrieval. Multi-hop reasoning datasets also provide insights into LLMs' internal mechanism~\cite{shortcut}, underscoring the importance of studying this ability.

Despite growing interest in reasoning, multi-hop reasoning of large audio-language models (LALMs) based on speech and audio remains unexplored. Existing benchmarks assess general speech and audio-processing performance~\cite{dynamicsuperb, airbench}, conversational abilities~\cite{styletalk, sdeval}, and fairness/bias~\cite{listenspeakfairly, spokenstereoset}, while studies on LALMs' fundamental abilities~\cite{ghosh2024compa, mmau, kuan24_interspeech} remain limited and do not specifically address multi-hop reasoning.

To address this gap, we introduce \textbf{SAKURA} (\textbf{\underline{S}}peech and \textbf{\underline{A}}udio-based Question-answering Benchmar\textbf{\underline{k}} for M\textbf{\underline{u}}lti-hop \textbf{\underline{R}}e\textbf{\underline{a}}soning of Large Audio-Language Models), a novel benchmark for systematic evaluation of LALMs' multi-hop reasoning using speech and audio information. SAKURA has four tracks based on speech/audio attributes including speaker gender, spoken language, speaker emotion, and animal sound. Each track includes \textbf{single-hop} and \textbf{multi-hop} sub-tracks, both comprising 500 multiple-choice questions. Single-hop questions require direct perception of an attribute (e.g., \textit{``What is \textbf{the animal in the sound}?"}), whereas multi-hop questions demand further reasoning based on the extracted information (e.g., \textit{``Which of the listed features aligns best with \textbf{the physical appearance} of \textbf{the animal in the sound}?"}). This design compels models to recall and integrate the extracted auditory cues when tackling multi-hop questions. Figure~\ref{fig:overview} provides examples of single-hop and multi-hop reasoning paths. In total, SAKURA offers 4000 human-verified multiple-choice questions, providing not only a structured framework for evaluating LALMs’ speech/audio-based multi-hop reasoning but also a valuable resource for future research on their reasoning mechanisms.


Through evaluating various LALMs \cite{gong_ltuas, gama, salmonn, desta2, Qwen-Audio, Qwen2-Audio, gpt4o, gemini} on SAKURA, we find that LALMs struggle to recognize certain attributes and fail in multi-hop reasoning even when recognizing accurately. Moreover, while LALMs excel in reasoning based on text, \textbf{they fail to integrate latent speech/audio representations into the reasoning process}, contradicting expectations for end-to-end LALMs that should internally unify speech/audio understanding with LLM reasoning capabilities.

Overall, our contributions are: (1) introducing SAKURA, the first benchmark for systematically evaluating LALMs’ multi-hop reasoning, and (2) uncovering LALMs’ difficulties in integrating speech/audio for multi-hop reasoning. Relevant resources such as dataset, evaluation code, etc., can be found at \url{https://github.com/ckyang1124/SAKURA}.

\section{Dataset construction}
\label{section:dataset}

\subsection{Overview}



SAKURA evaluates LALMs' multi-hop reasoning with speech/audio information across \textbf{four} tracks, covering fundamental attributes: \textbf{speaker gender} (Gender), \textbf{spoken language} (Language), \textbf{speaker emotion} (Emotion), and \textbf{animal sounds} (Animal), chosen for their significance in speech/audio processing. Each track includes \textbf{single-hop} and \textbf{multi-hop} sub-tracks. Here, we define the single-hop reasoning of LALMs as the direct perception and extraction of speech/audio attributes, forming the foundation for speech/audio-based multi-hop reasoning.


In \textbf{single-hop} sub-tracks, models must extract an attribute directly from a speech/audio input (e.g., identifying the animal sound). In \textbf{multi-hop} sub-tracks, models must use the extracted information for further reasoning (e.g., determining a physical trait that matches the identified animal). Each sub-track contains 500 multiple-choice questions, each consisting of (1) a speech/audio input, (2) a textual question with a fixed number of choices, and (3) a golden answer. The multiple-choice format was adopted to facilitate standardized assessment and comparison. Additionally, speech/audio inputs are shared across sub-tracks within each track, enabling controlled evaluation of reasoning relative to perception accuracy. Figure~\ref{fig:overview} provides a data example, and Table~\ref{tab:statistics} shows key dataset statistics, showcasing its diversity without strong biases toward any particular ranges.
\vspace{-10pt}
\begin{table}[ht]\small
\setlength\tabcolsep{3pt}
\renewcommand{\arraystretch}{0.02}

\caption{Statistics of SAKURA. ``Avg", ``Std", ``Min", and ``Max" represent the average, standard deviation, minimum, and maximum of the measured quantities.}
\centering
\vspace{-5pt}
\begin{tabular}{c|c|c|c|c}

\toprule
 & Avg & Std & Min & Max \\
\midrule

 \makecell[c]{Speech/audio duration (s)} & 4.79 & 2.23 & 0.22 & 20.78 \\

\midrule
 
 \makecell[c]{Instruction length (words)} & 31.32 & 12.57 & 8 & 66 \\

 \bottomrule

\end{tabular}

\label{tab:statistics}
\end{table}

\vspace{-8pt}
\subsection{Speech/audio sources}
We describe the speech/audio data sources for each track:
\begin{itemize}
    \item \textbf{Gender}: we randomly selected 500 samples from the testing split of the English subset of Common Voice 17.0~\cite{commonvoice:2020}, which contains over 20k hours of validated human speech in 124 languages. We balance the genders of the speakers to mitigate potential biases.
    \item \textbf{Language}: we drew 500 samples from the testing splits of eight languages\footnote{English/German/Spanish/French/Italian/Chinese/Japanese/Korean} in Common Voice 17.0, while balancing both the language distribution and the speaker genders.
    \item \textbf{Emotion}: we randomly chose 500 samples from CREMA-D~\cite{cao2014crema} and MELD~\cite{meld1, meld2}, both widely used for speech emotion recognition. Samples were evenly distributed across five emotions\footnote{Happy / Disgust / Sad / Fear / Angry} while ensuring gender balance among speakers.
    \item \textbf{Animal}: we collected 500 audio samples from ESC-50~\cite{esc50} and the dataset curated by {\c{S}}a{\c{s}}maz et al.~\cite{animalsound}. ESC-50 contains 5-second environmental sounds categorized into 50 categories, from which we selected samples from nine animal categories\footnote{Dog / Cat / Pig / Cow / Frog / Hen / Rooster / Sheep / Crow}. We included additional samples from {\c{S}}a{\c{s}}maz et al., selecting only those matching the same nine categories.
\end{itemize}





\begin{table*}[ht]\small 
\setlength\tabcolsep{3pt} 
\renewcommand{\arraystretch}{0.9}
\caption{Accuracies (\%) and 95\% confidence intervals of the baselines on SAKURA. ``Single" and ``Multi" denote the single-hop and multi-hop sub-tracks. The best and the second-best performances among open-source and proprietary LALMs are marked in bold and underlined. Model sizes (except proprietary ones) are provided, with cascaded models showing the sum of involved modules.}

\centering

\resizebox{0.96\linewidth}{!}{
\begin{tabular}{l|c|cc|cc|cc|cc|cc} 
  \toprule
  & \multirow{2.6}{*}{\makecell[c]{\textbf{Size}\\\textbf{(B)}}} & \multicolumn{2}{c|}{\textbf{Gender}} & \multicolumn{2}{c|}{\textbf{Language}} & \multicolumn{2}{c|}{\textbf{Emotion}} & \multicolumn{2}{c|}{\textbf{Animal}} & \multicolumn{2}{c}{\textbf{Average}} \\ 
  \cmidrule(lr){3-12}
   &  & Single & Multi & Single & Multi & Single & Multi & Single & Multi & Single & Multi \\ 
  \midrule
  \multicolumn{12}{c}{\textbf{Open-source LALMs}}\\
  \midrule
  LTU-AS & 7 & 52.4 \smaller$\pm$ 4.4&19.6 \smaller$\pm$ 3.5&16.8 \smaller$\pm$ 3.3&11.4 \smaller$\pm$ 2.8&28.6 \smaller$\pm$ 4.0&19.6 \smaller$\pm$ 3.5&65.6 \smaller$\pm$ 4.2&21.8 \smaller$\pm$ 3.6&40.9 \smaller$\pm$ 4.3&18.1 \smaller$\pm$ 3.4\\
  
  GAMA-IT&7&76.4 \smaller$\pm$ 3.7 &39.8 \smaller$\pm$ 4.3 &5.6 \smaller$\pm$ 2.0 &19.4 \smaller$\pm$ 3.5 &5.6 \smaller$\pm$ 2.0 &24.2 \smaller$\pm$ 3.8 &85.2 \smaller$\pm$ 3.1 &51.4 \smaller$\pm$ 4.4 &43.2 \smaller$\pm$ 4.3 &33.7 \smaller$\pm$ 4.1\\

  SALMONN & 7.5 & 59.8 \smaller$\pm$ 4.3 &\underline{48.6} \smaller$\pm$ 4.4 &21.8 \smaller$\pm$ 3.6 &29.6 \smaller$\pm$ 4.0 &19.8 \smaller$\pm$ 3.5 &28.2 \smaller$\pm$ 3.9 &68.6 \smaller$\pm$ 4.1 &34.6 \smaller$\pm$ 4.2 &42.5 \smaller$\pm$ 4.3 &35.3 \smaller$\pm$ 4.2 \\
  
  DeSTA2 & 8.3 & \textbf{88.4} \smaller$\pm$ 2.8 &\textbf{85.2} \smaller$\pm$ 3.1 &94.2 \smaller$\pm$ 2.0 &75.4 \smaller$\pm$ 3.8 &34.8 \smaller$\pm$ 4.2 &36.4 \smaller$\pm$ 4.2 &34.4 \smaller$\pm$ 4.2 &31.2 \smaller$\pm$ 4.1 &63.0 \smaller$\pm$ 4.2 &\textbf{57.1} \smaller$\pm$ 4.3 \\
    
  
  
  Qwen-Audio-Chat & 8.4 &49.6 \smaller$\pm$ 4.4 &43.8 \smaller$\pm$ 4.3 &87.6 \smaller$\pm$ 2.9 &40.6 \smaller$\pm$ 4.3 &\underline{63.2} \smaller$\pm$ 4.2 &\underline{37.0} \smaller$\pm$ 4.2 &\textbf{92.2} \smaller$\pm$ 2.4 &\textbf{66.0} \smaller$\pm$ 4.2 &\underline{73.2} \smaller$\pm$ 3.9 &46.9 \smaller$\pm$ 4.4 \\

  Qwen2-Audio-Instruct & 8.4 &\underline{88.0} \smaller$\pm$ 2.8 &47.2 \smaller$\pm$ 4.4 &83.8 \smaller$\pm$ 3.2 &48.0 \smaller$\pm$ 4.4 &\textbf{64.2} \smaller$\pm$ 4.2 &\textbf{39.8} \smaller$\pm$ 4.3 &\underline{88.8} \smaller$\pm$ 2.8 &\underline{61.4} \smaller$\pm$ 4.3 &\textbf{81.2} \smaller$\pm$ 3.4 &49.1 \smaller$\pm$ 4.4 \\
  
  
  
  \midrule
  \multicolumn{12}{c}{\textbf{Proprietary LALMs}}\\
  \midrule
  
  GPT-4o Audio & - & - & - &95.2 \smaller$\pm$ 1.9 &\underline{83.6} \smaller$\pm$ 3.2 &38.2 \smaller$\pm$ 4.3 &23.8 \smaller$\pm$ 3.7 &80.6 \smaller$\pm$ 3.5 &55.4 \smaller$\pm$ 4.4 &71.3 \smaller$\pm$ 4.0 &\underline{54.3} \smaller$\pm$ 4.4\\
  
  Gemini-1.5-flash & - & 77.0 \smaller$\pm$ 3.7 &24.2 \smaller$\pm$ 3.8 &\textbf{98.2} \smaller$\pm$ 1.2 &79.8 \smaller$\pm$ 3.5 &24.6 \smaller$\pm$ 3.8 &19.4 \smaller$\pm$ 3.5 &27.2 \smaller$\pm$ 3.9 &16.2 \smaller$\pm$ 3.2 &56.8 \smaller$\pm$ 4.3 &34.9 \smaller$\pm$ 4.2 \\
  Gemini-1.5-pro & - & 74.0 \smaller$\pm$ 3.8 &43.4 \smaller$\pm$ 4.3 &\underline{97.2} \smaller$\pm$ 1.4 &\textbf{90.6} \smaller$\pm$ 2.6 &39.2 \smaller$\pm$ 4.3 &24.0 \smaller$\pm$ 3.7 &42.0 \smaller$\pm$ 4.3 &28.6 \smaller$\pm$ 4.0 &63.1 \smaller$\pm$ 4.2 &46.6 \smaller$\pm$ 4.4\\
  
  \midrule
  \multicolumn{12}{c}{\textbf{Cascaded Systems}}\\
  \midrule
  ASR+LLM & 9.5  & 24.2 \smaller$\pm$ 3.8 &32.2 \smaller$\pm$ 4.1 &93.6 \smaller$\pm$ 2.1 &82.4 \smaller$\pm$ 3.3 &21.4 \smaller$\pm$ 3.6 &30.6 \smaller$\pm$ 4.0 &30.8 \smaller$\pm$ 4.0 &27.6 \smaller$\pm$ 3.9 &42.5 \smaller$\pm$ 4.3 &43.2 \smaller$\pm$ 4.3 \\
  
  
  ASR+AAC+LLM & 17.9 & 85.0 \smaller$\pm$ 3.1 &79.6 \smaller$\pm$ 3.5 &93.4 \smaller$\pm$ 2.2 &88.8 \smaller$\pm$ 2.8 &60.0 \smaller$\pm$ 4.3 &51.4 \smaller$\pm$ 4.4 &78.0 \smaller$\pm$ 3.6 &78.4 \smaller$\pm$ 3.6 &79.1 \smaller$\pm$ 3.6 &74.5 \smaller$\pm$ 3.8 \\
  
  
  
  \midrule
  \multicolumn{12}{c}{\textbf{Random Baseline}}\\
  \midrule
  Chance level & - & 50.0 & 50.0 & 25.0 & 25.0 & 25.0 & 25.0 & 25.0 & 25.0 & 31.3 & 31.3 \\
  \bottomrule

\end{tabular}
}
\label{tab:main_result}

\vspace{-10pt}
\end{table*}

\subsection{Textual question-answer pair generation}
Figure~\ref{fig:overview} illustrates our question-answer data generation pipeline. We used GPT-4o\footnote{gpt-4o-2024-11-20, \url{https://openai.com}} to generate question-answer pairs. In the \textbf{single-hop} sub-tracks, questions directly queried target attributes. To enhance diversity, we used GPT-4o to paraphrase manually crafted questions, such as \textit{``What is the animal in the sound?"}, producing multiple variations with differing phrasings and lengths. These paraphrased questions were then paired with speech/audio data, with the associated attribute label as the golden answer and distractors selected from alternative labels.


The generation process for the \textbf{multi-hop} sub-tracks was more complex. For each track, we used GPT-4o to generate several ``tested topics" and three sample questions per topic to clarify their meaning. A topic is a concept that multi-hop questions can target based on the track's attribute labels (e.g., ``physical traits" in the Animal track). To ensure quality and diversity, we manually filtered the topics using these criteria:
\begin{enumerate}
    
    \item \textbf{Relevance and Groundedness}: Topics must be relevant to the track's attribute and supported by the corresponding information. For example, topics like ``determining the age of the speaker" in the Gender track or ``determining the religious beliefs of the speaker" in the Language track were excluded.
    
    \item \textbf{Objectivity}: Topics must be objective and unbiased. For example, ``the association of animals and human personality" was excluded due to cultural ambiguity.

    \item \textbf{Uniqueness}: Topics should be distinct. For example, only one of ``feeding habits" and ``food sources" could be included.
\end{enumerate}



For each topic, GPT-4o generated diverse questions along with golden answer candidates and distractors for all attribute labels, forming a choice pool for pairing with speech/audio data. For example, for a question on ``physical traits" with ``cat" as the attribute label, GPT-4o might generate ``claws" as a golden answer candidate and ``tentacles" and ``hooves" as distractors.

To ensure data quality, human annotators rigorously verified the correctness of golden answer candidates and the inaccuracy of distractors by cross-referencing reliable sources (e.g., Wikipedia) based on the questions and attribute labels. Each golden answer candidate and distractor was reviewed by at least three annotators and accepted only if all annotators agreed, ensuring the quality. Once validated, questions were paired with corresponding speech/audio data, and the verified choice pools were used to sample the appropriate golden answer and distractors based on the attribute label of the paired speech/audio input. 

Before finalizing the instructions, the choice order, including the golden answer and distractors, was shuffled to ensure the golden answer appeared equally frequent across positions (e.g., (a), (b), etc.), mitigating potential positional biases in LALMs (i.e., favoring a certain option, such as ``(a)", irrespective of content). The instruction was constructed by concatenating the question with the shuffled choices, with option prefixes (e.g., ``(a)", ``(b)") inserted. The paired speech/audio inputs, instructions, and golden answers were included in SAKURA.



\begin{table*}[ht]\small 
\setlength\tabcolsep{3pt} 
\renewcommand{\arraystretch}{0.8}

\caption{Accuracies (\%) and 95\% confidence intervals of open-source and proprietary LALMs on multi-hop sub-tracks, considering only instances where the corresponding single-hop questions were correctly answered. Results under speech/audio-based (“S/A”) and text-based (“Text”) reasoning settings are reported. The better performances among the two settings are marked in bold.}

\centering

\resizebox{0.97\linewidth}{!}{
\begin{tabular}{l|cc|cc|cc|cc|cc} 
  \toprule
   & \multicolumn{2}{c|}{\textbf{Gender}} & \multicolumn{2}{c|}{\textbf{Language}} & \multicolumn{2}{c|}{\textbf{Emotion}} & \multicolumn{2}{c|}{\textbf{Animal}} & \multicolumn{2}{c}{\textbf{Average}}
  \\ 
  \cmidrule(lr){2-11}
   &  S/A & Text & S/A & Text & S/A & Text & S/A & Text & S/A & Text \\
  \midrule
  
  LTU-AS & 19.5 \smaller$\pm$ 3.5 &\textbf{23.3} \smaller$\pm$ 3.7 &20.2 \smaller$\pm$ 3.5 &\textbf{42.9} \smaller$\pm$ 4.3 &28.7 \smaller$\pm$ 4.0 &\textbf{32.2} \smaller$\pm$ 4.1 &\textbf{22.9} \smaller$\pm$ 3.7 &18.3 \smaller$\pm$ 3.4 &22.8 \smaller$\pm$ 3.7 &\textbf{32.8} \smaller$\pm$ 4.1 \\
  
  


  GAMA-IT & 44.5 \smaller$\pm$ 4.4 &\textbf{69.6} \smaller$\pm$ 4.0 &28.6 \smaller$\pm$ 4.0 &\textbf{64.3} \smaller$\pm$ 4.2 &35.7 \smaller$\pm$ 4.2 &\textbf{64.3} \smaller$\pm$ 4.2 &53.5 \smaller$\pm$ 4.4 &\textbf{74.2} \smaller$\pm$ 3.8 &40.6 \smaller$\pm$ 4.3 &\textbf{68.1} \smaller$\pm$ 4.1 \\
  
  
  SALMONN  & 51.8 \smaller$\pm$ 4.4 &\textbf{76.6} \smaller$\pm$ 3.7 &48.6 \smaller$\pm$ 4.4 &\textbf{83.5} \smaller$\pm$ 3.3 &38.4 \smaller$\pm$ 4.3 &\textbf{56.6} \smaller$\pm$ 4.3 &34.1 \smaller$\pm$ 4.2 &\textbf{58.0} \smaller$\pm$ 4.3 &43.2 \smaller$\pm$ 4.3 &\textbf{68.7} \smaller$\pm$ 4.1 \\
  

  DeSTA2 & 90.0 \smaller$\pm$ 2.6 &\textbf{98.6} \smaller$\pm$ 1.0 &78.6 \smaller$\pm$ 3.6 &\textbf{93.6} \smaller$\pm$ 2.1 &51.1 \smaller$\pm$ 4.4 &\textbf{94.2} \smaller$\pm$ 2.0 &43.6 \smaller$\pm$ 4.3 &\textbf{86.6} \smaller$\pm$ 3.0 &65.8 \smaller$\pm$ 4.2 &\textbf{93.3} \smaller$\pm$ 2.2 \\
  
  Qwen-Audio-Chat & 40.3 \smaller$\pm$ 4.3 &\textbf{55.6} \smaller$\pm$ 4.4 &40.2 \smaller$\pm$ 4.3 &\textbf{73.3} \smaller$\pm$ 3.9 &44.9 \smaller$\pm$ 4.4 &\textbf{69.0} \smaller$\pm$ 4.1 &68.3 \smaller$\pm$ 4.1 &\textbf{80.0} \smaller$\pm$ 3.5 &48.4 \smaller$\pm$ 4.4 &\textbf{69.5} \smaller$\pm$ 4.0 \\
  
  
  Qwen2-Audio-Instruct & 49.3 \smaller$\pm$ 4.4 &\textbf{72.5} \smaller$\pm$ 3.9 &50.4 \smaller$\pm$ 4.4 &\textbf{74.7} \smaller$\pm$ 3.8 &45.5 \smaller$\pm$ 4.4 &\textbf{74.8} \smaller$\pm$ 3.8 &62.2 \smaller$\pm$ 4.3 &\textbf{75.7} \smaller$\pm$ 3.8 &51.8 \smaller$\pm$ 4.4 &\textbf{74.4} \smaller$\pm$ 3.8 \\
  
  
  \midrule
  \midrule
  GPT-4o Audio & - & - & 86.5 \smaller$\pm$ 3.0 &\textbf{91.6} \smaller$\pm$ 2.4 &38.2 \smaller$\pm$ 4.3 &\textbf{73.3} \smaller$\pm$ 3.9 &63.0 \smaller$\pm$ 4.2 &\textbf{68.5} \smaller$\pm$ 4.1 &62.6 \smaller$\pm$ 4.2 &\textbf{77.8} \smaller$\pm$ 3.6 \\
  
  Gemini-1.5-flash &27.5 \smaller$\pm$ 3.9 &\textbf{78.7} \smaller$\pm$ 3.6 &80.9 \smaller$\pm$ 3.4 &\textbf{95.5} \smaller$\pm$ 1.8 &37.4 \smaller$\pm$ 4.2 &\textbf{85.4} \smaller$\pm$ 3.1 &20.6 \smaller$\pm$ 3.5 &\textbf{71.3} \smaller$\pm$ 4.0 &41.6 \smaller$\pm$ 4.3 &\textbf{82.7} \smaller$\pm$ 3.3 \\

  Gemini-1.5-pro & 48.1 \smaller$\pm$ 4.4 &\textbf{91.3} \smaller$\pm$ 2.5&91.6 \smaller$\pm$ 2.4 &\textbf{97.7} \smaller$\pm$ 1.3 &40.3 \smaller$\pm$ 4.3 &\textbf{79.1} \smaller$\pm$ 3.6 &34.8 \smaller$\pm$ 4.2 &\textbf{69.5} \smaller$\pm$ 4.0 &53.7 \smaller$\pm$ 4.4 &\textbf{84.4} \smaller$\pm$ 3.2 \\
  
  
  \bottomrule

\end{tabular}
}
\label{tab:text-based}
\vspace{-8pt}
\end{table*}
\section{Experimental setups}
\subsection{Evaluation metrics}
Since SAKURA comprises multiple-choice questions, accuracy is a natural metric. However, because LALMs sometimes generate descriptive responses instead of explicitly selecting a choice, we adopt an LLM-as-a-judge~\cite{llmjudge} approach following prior works~\cite{airbench, dynamicsuperb}.

Specifically, we employ GPT-4o\footnote{gpt-4o-2024-11-20, \url{https://openai.com}} as the evaluator. During evaluation, the evaluator is provided with well-defined criteria, the original instruction, the golden answer, and the response from the evaluated model. It then determines whether the response aligns with the golden answer, following these rules:
\begin{enumerate}
    \item Each question in SAKURA has exactly one correct answer. If a model fails to select \textbf{one and only one} choice from the given choices, either by selecting multiple choices or failing to explicitly choose any, it should be marked as \textbf{incorrect}.

    \item The evaluator must assess the alignment between the golden answer and the model’s response, providing an explanation to enhance the evaluation transparency.

    \item The final judgment should be summarized as a binary ``correct/incorrect" label to facilitate post-processing.
\end{enumerate}

We incorporate in-context examples to enhance evaluation quality. To verify the evaluator’s reliability, we conducted human verification on its judgments on 200 randomly selected samples, finding 99.5\% agreement with human annotations. This confirms the robustness of our LLM-based evaluation. Final accuracy is computed based on the judgments.

\subsection{Baselines}
We incorporated models from three categories: open-source LALMs, proprietary LALMs, and cascaded systems. For all baselines, we applied greedy decoding without system prompts. 

Open-source LALMs involved models of comparable sizes, including LTU-AS\footnote{\url{https://github.com/YuanGongND/ltu}}~\cite{gong_ltuas}, 
GAMA-IT\footnote{\url{https://github.com/Sreyan88/GAMA}}~\cite{gama}, SALMONN\footnote{\url{https://github.com/bytedance/SALMONN}}~\cite{salmonn}, Qwen-Audio-Chat\footnote{\url{https://github.com/QwenLM/Qwen-Audio}}~\cite{Qwen-Audio}, Qwen2-Audio-Instruct\footnote{\url{https://github.com/QwenLM/Qwen2-Audio}}~\cite{Qwen2-Audio}, and DeSTA2\footnote{\url{https://github.com/kehanlu/DeSTA2}}~\cite{desta2}. Proprietary LALMs included GPT-4o Audio\footnote{gpt-4o-audio-preview-2024-12-17, \url{https://openai.com}}~\cite{gpt4o}, Gemini-1.5-flash~\cite{gemini}, and Gemini-1.5-pro\footnote{gemini-1.5-flash-002 and gemini-1.5-pro-002, \par\url{https://deepmind.google/technologies/gemini/}}~\cite{gemini}.

Cascaded systems comprised three components: a speech recognition (ASR) module, an audio captioning (AAC) module, and a text-based LLM, forming two variants: (1) ``ASR+LLM," where the LLM answered questions based only on ASR transcriptions, and (2) ``ASR+AAC+LLM," where audio captions enriched the input to the LLM. Specifically, we used Whisper-large-v3~\cite{whisper} for ASR, Qwen2-Audio-Instruct for AAC, and LLaMA-3.1-8B-Instruct~\cite{llama31} as the LLM.

Experiments on models except for proprietary LALMs took 68 V100 GPU hours, while those on proprietary LALMs took 15.5 hours. The full LLM-based evaluation spanned 62 hours.

\section{Results}
\subsection{Main results on SAKURA}
\label{sec:main_results}

Table~\ref{tab:main_result} shows the baseline performances on SAKURA, with chance level included for reference. GPT-4o Audio’s results on the Gender track are omitted as it refuses to answer gender-related questions due to post-training constraints, consistent with prior findings~\cite{lin2025preliminaryexplorationgpt4ovoice}.


In the \textit{single-hop} sub-tracks, which assess perception (information extraction) ability, Qwen2-Audio-Instruct achieves the highest average accuracy. However, no model consistently outperforms across all tracks, with each exhibiting distinct blind spots, indicating weaknesses in processing certain speech/audio attributes. For instance, DeSTA2 and Gemini-1.5-flash struggle on the Emotion and Animal tracks despite high accuracy elsewhere, while GAMA-IT and SALMONN perform worse than chance on the Language and Emotion tracks. Notably, despite being trained on emotion-related tasks, most LALMs struggle with the Emotion track, likely due to the inherent subtlety of emotional cues that require more nuanced perception. \textbf{Our findings highlight the need for improving LALMs’ fundamental perception capabilities}.


In the \textit{multi-hop} sub-tracks, accuracy drops drastically compared to single-hop ones, even when models exhibit good perception. For example, although Qwen-Audio-Chat and Qwen2-Audio-Instruct excel at recognizing emotion and animal sounds, and DeSTA2 and Gemini-1.5-flash identify the languages well, they all face substantial declines in multi-hop reasoning. This sharp contrast suggests that while LALMs may extract correct information, \textbf{they struggle to reason based on that information}, exposing a fundamental limitation in their speech/audio-based multi-hop reasoning ability.


Comparing model groups, the best LALMs in both single-hop and multi-hop sub-tracks are generally open-source, suggesting that proprietary models do not always have a competitive edge. Their advantage appears only in the Language track, likely due to larger and more diverse pre-training datasets, but it diminishes in other tracks. Additionally, ASR+AAC+LLM achieves higher average accuracy than most LALMs on both sub-tracks, showing that \textbf{current LALMs still fall short of cascaded approaches}, as observed in prior works~\cite{kuan2024speech, mmau}.


\subsection{Fail to reason, or fail to reason with speech and audio?}
\label{sec:text-based}

To further investigate why LALMs struggle with multi-hop reasoning, we compared their performances under two conditions: (1) \textbf{speech/audio-based} reasoning, where models process raw speech/audio inputs, and (2) \textbf{text-based} reasoning, where these inputs are replaced by captions summarizing the relevant attribute information (e.g., “The animal making the sound is cat.”) alongside the original instructions (see Figure~\ref{fig:overview}). This comparison examines how LALMs reason using encoded auditory representations compared to an approximate textual equivalent summarizing relevant information. For each model, we filtered the multi-hop sub-tracks to include only instances where it correctly answered the corresponding single-hop questions. This ensured that failures in multi-hop reasoning stemmed from reasoning limitations rather than misperception of attributes.


As shown in Table~\ref{tab:text-based}, most LALMs struggle with speech/audio-based reasoning despite \textbf{correctly recognizing attributes}, yet perform \textbf{significantly better when similar information is provided as text}. For example, DeSTA2 exceeds 90\% accuracy in the text-based setting, demonstrating strong reasoning ability when leveraging explicitly provided textual information alongside stored knowledge. Since the only difference between these conditions is the modality used to present attribute information, this discrepancy suggests that while LALMs possess basic reasoning abilities, \textbf{their reasoning remains predominantly text-driven, revealing a lack of true multimodal integration}. Even when extracting accurate information, LALMs \textbf{fail to incorporate latent speech/audio representations into the reasoning process}, contradicting expectations for end-to-end models to unify speech/audio understanding with reasoning. Our findings highlight the urgent need for improved multimodal reasoning capabilities.

\section{Conclusion, limitations, and future work}



We introduce SAKURA, the first benchmark for systematically evaluating LALMs' multi-hop reasoning with speech and audio. Our findings show that LALMs struggle to recognize certain speech and audio attributes, exhibiting perception blind spots. Even when accurately extracting relevant information, they fall short in speech/audio-based multi-hop reasoning despite good text-based reasoning skills, highlighting the need for better integration of multimodal information into their reasoning process.


SAKURA has certain limitations that we aim to address in future work. As an initial study, it covers only four speech/audio attributes and does not specifically account for acoustic variations (e.g., background noise, speaking rate) that may affect model performance. Expanding both the range of attributes and the diversity of acoustic conditions is a key future direction for assessing model robustness. Additionally, while SAKURA currently employs a multiple-choice evaluation for text-generating LALMs, future iterations will investigate speech-generating models~\cite{llama-omni, moshi, yang2024building} and open-ended tasks for deeper insights.

\section{Acknowledgement}
We acknowledge the computational and storage support provided by the National Center for High-performance Computing (NCHC) of the National Applied Research Laboratories (NARLabs) in Taiwan.

\bibliographystyle{IEEEtran}
\bibliography{mybib}

\end{document}